\documentclass{ws-p8-50x6-00}

\newcommand{\ovl}{\overline}
\newcommand{\ul}{\underline}
\newcommand{\Pslash}{\kern 0.2 em P\kern -0.56em \raisebox{0.3ex}{/}}
\newcommand{\pslash}{\kern 0.2 em p\kern -0.4em /}
\newcommand{\kslash}{\kern 0.2 em k\kern -0.45em /}
\newcommand{\Sslash}{\kern 0.2 em S\kern -0.56em \raisebox{0.3ex}{/}}
\newcommand{\Mslash}{\kern 0.2 em M\kern -0.70em \raisebox{0.3ex}{/}}
\newcommand{\sumint}{\kern 0.2 em {\textstyle\sum} \kern -1.1 em \int_X}
\newcommand{\g}{\gamma}
\newcommand{\sig}{\sigma}
\newcommand{\eps}{\epsilon}
\newcommand{\dg}{\dagger}
\newcommand{\xb}{x_{\scriptscriptstyle B}}
\newcommand{\sT}{{\scriptscriptstyle T}}
\newcommand{\newangle}{{<\kern -0.3 em{\scriptscriptstyle )}}}
\newcommand{\lsim}{\raisebox{-4pt}{$\,\stackrel{\textstyle <}{\sim}\,$}}
\newcommand{\nn}{\nonumber}

\begin{document}

\title{
Interference Fragmentation Functions\\
in Deep Inelastic Scattering}

\author{S.~Boffi, \underline{R.~Jakob}, M.~Radici}

\address{Universit\`a di Pavia e Istituto Nazionale di Fisica Nucleare, 
Sezione di Pavia, Pavia, Italy}

\author{A.~Bianconi}

\address{Universit\`a di Brescia e Istituto Nazionale di Fisica Nucleare, 
Sezione di Pavia, Pavia, Italy}  

%%%%%%%%%%%%%%%%%%%%%%%%%%%%%%%%%%%%%%%%%%%%%%%%%%%%%%%%%%%%%%

\maketitle

\abstracts{We give a general overview of the r\^ole and interpretation 
of fragmentation functions in hard processes. Transverse momentum dependence
gives rise to time-reversal odd fragmentation functions contributing at 
leading order. Final state interactions are necessary to have non-zero 
time reversal odd functions. We present a model calculation for the 
special case of interference fragmentation
functions in two-hadron inclusive lepton hadron scattering.\\[2mm]
{\footnotesize Talk presented at the 
2nd Int.~Conference on "Perspectives in Hadronic Physics", 10-14 May 1999, 
Miramare - Trieste, Italy}}

%%%%%%%%%%%%%%%%%%%%%%%%%%%%
\noindent
\hspace*{\fill}\raisebox{89mm}[-89mm]{\small hep-ph/9907374}
\vspace*{-4mm}

\section{Physics motivation}

Hard processes like deep-inelastic lepton-hadron scattering (DIS), the
electron/positron annihilation ($e^+e^-$), or lepton-pair poduction 
in hadron-hadron scattering probe the {\em internal structure of hadrons 
in terms of quark and gluon degrees of freedom}.

The information obtained on the hadronic structure is encoded in 
{\em distribution functions} (DF) and {\em fragmentation functions} (FF).
Those functions taken from one experiment can be 
used without change to predict the results of other hard processes. The 
property of independence from the particular reaction, the 
so-called {\em universality}, is assured by the existence of factorization 
theorems for some of the hard processes\cite{col89}, whereas for others 
factorization is used as a plausible assumption.

Leading twist distribution and fragmentation functions (i.e.\ those which 
contribute to leading order in an expansion of inverse powers of the hard 
scale $Q$ in the process under consideration) have a 
simple {\em probabilistic interpretation} in the context of lightcone
quantization.

The parton model provides us with simple relations between DF/FF and
the physical observables of the processes, like {\em cross sections} or
{\em structure functions}.

%%%%%%%%%%%%%%%%%%%%%%%%%%%%%%%%%%%%%%%%%%%%
\section{Quark-quark correlation functions\\
distribution/fragmentation functions}
We illustrate the formal definitions of distribution/fragmentation 
functions and their relations to observables as given by the 
(na\"{\i}ve, i.e.\ not QCD corrected) parton model with two 
examples:
\begin{itemize} 
\item
The hadron tensor for {\em inclusive deep inelastic lepton-hadron scattering} 
is calculated in leading order from the ``handbag'' diagram\\[-6mm]
\begin{center} 
\parbox[b]{36mm}{\includegraphics[width=36mm]{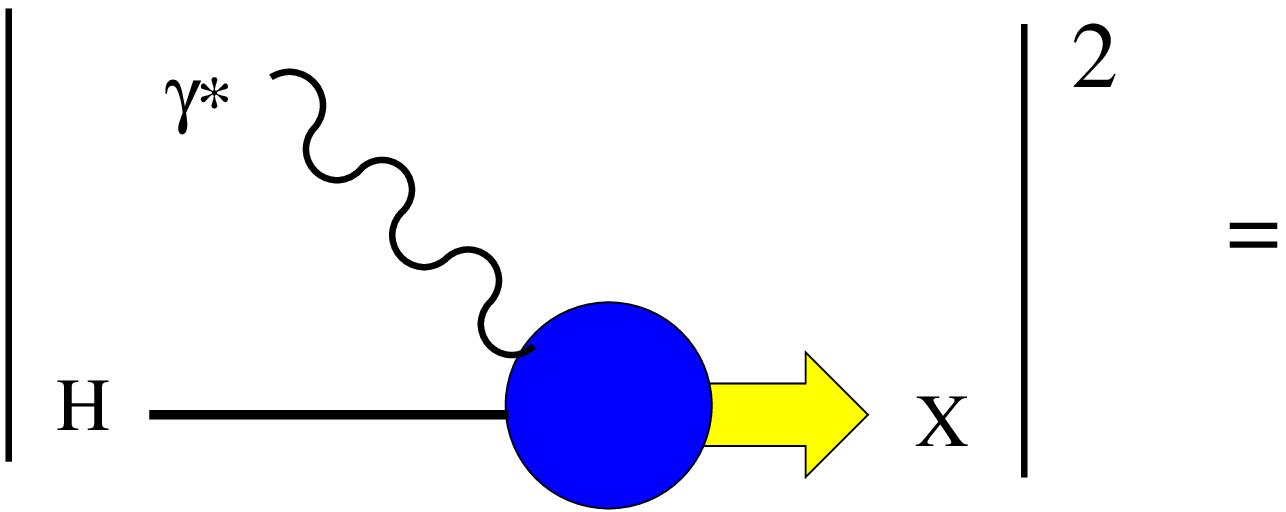}\\[2mm]}
\qquad
\parbox[b]{44mm}{\includegraphics[width=44mm]{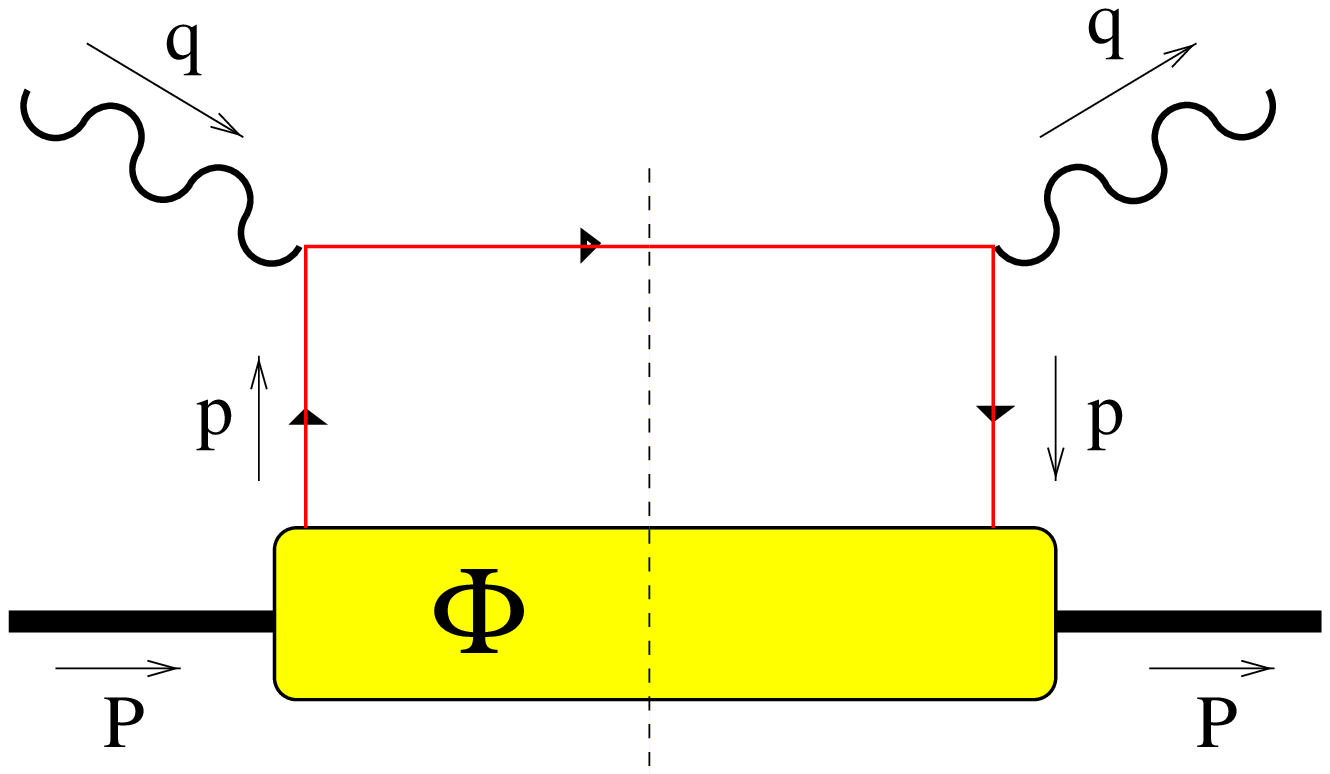}}
\end{center}
\vspace{-2mm}
The famous parton model relation between the structure function 
$F_1$ (and $F_2$ via the Callan-Gross relation) and the DF $f_1$ is
\begin{equation}
2F_1(x_B)={F_2(x_B)\over x_B}=\sum_a e_a^2\,f_1^a(x_B)
\end{equation}
with $x_B=Q^2/(2P\cdot q)$ and summation over flavors $a$. A 
formal definition\cite{SoperCollins,Jaffe83} 
of $f_1$ is obtained from a quark-quark correlation function
\begin{equation} 
\Phi_{ij}(p;P,S)=\int\frac{d^{\,4}x}{(2\pi)^4}\;e^{ip\cdot x}\;
\langle P,S|\overline\psi_j(0)\;\psi_i(x)|P,S\rangle
\end{equation} 
as (partially integrated) Dirac projection in the form
\begin{equation} 
f_1(x) = 
\left.\frac{1}{2}\int dp^-\,d^{\,2}\vec p_\sT\;{\rm Tr}(\Phi\;\gamma^+) 
\right|_{p^+ = x P^+}
\end{equation} 

\item
The leading order diagram for 
{\em one-hadron inclusive electron/positron annihilation} is

\begin{center} 
\parbox[b]{36mm}{\includegraphics[width=36mm]{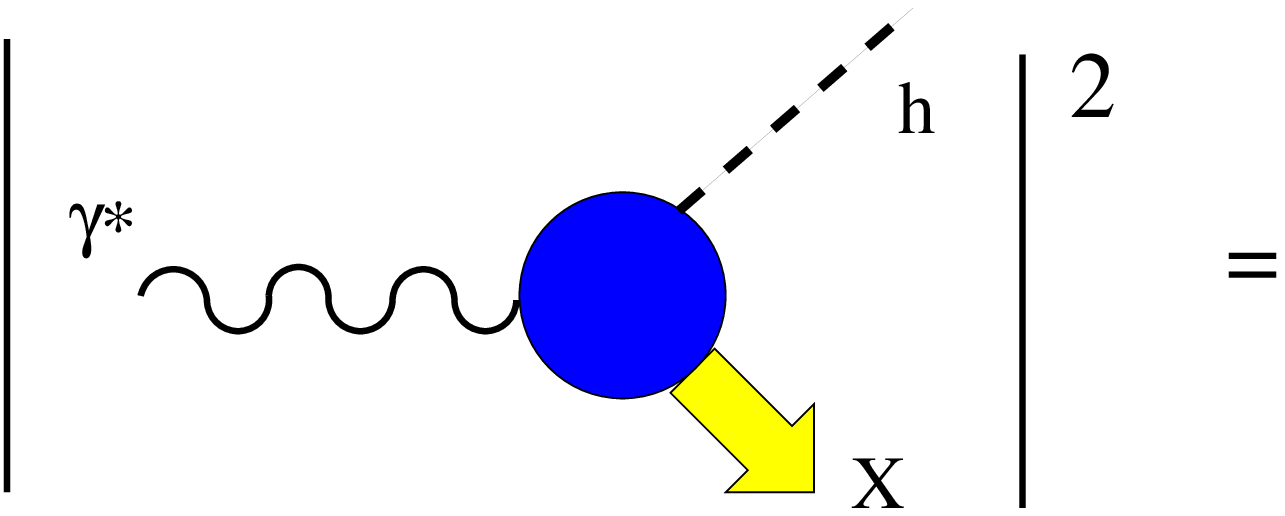}\\[2mm]}
\qquad
\parbox[b]{44mm}{\includegraphics[width=44mm]{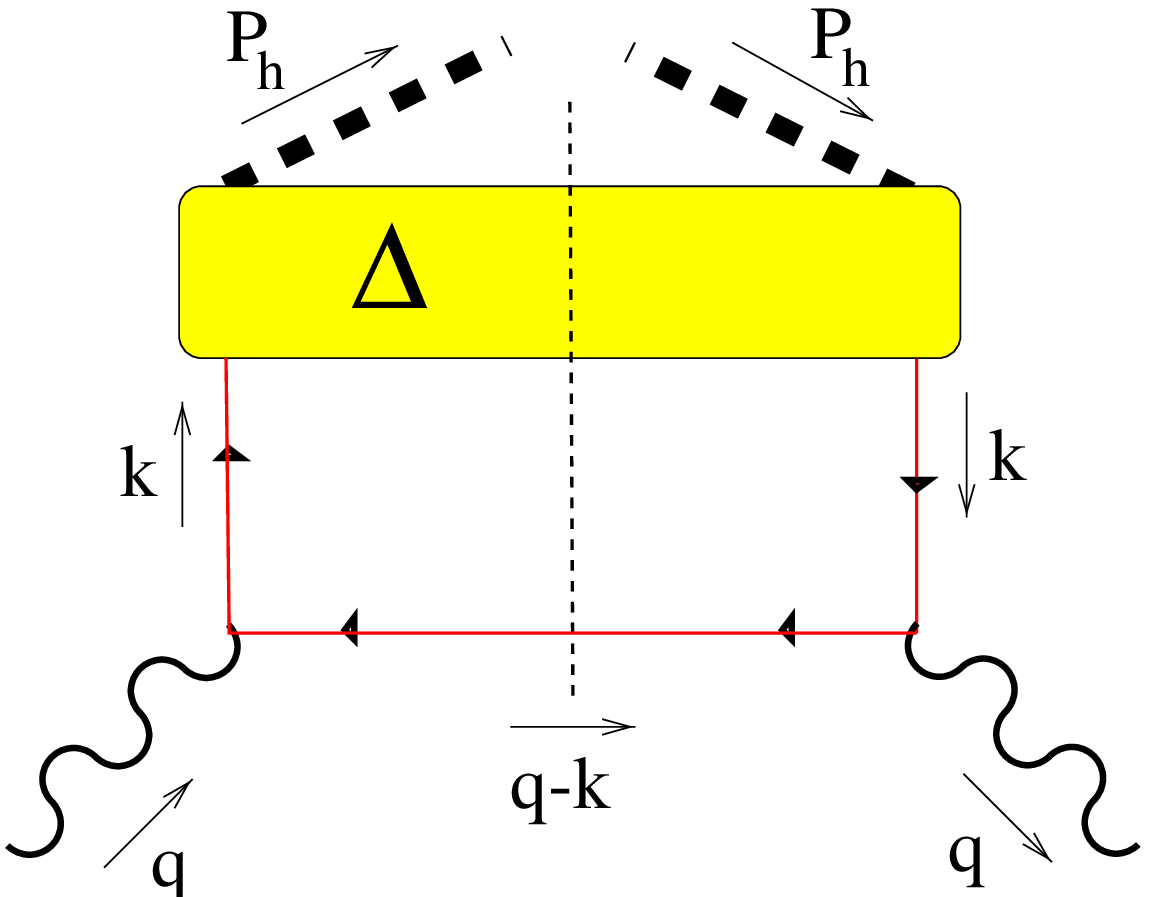}}
\end{center}
The parton model relation between the cross section, and the FF $D_1$ reads
\begin{equation} 
{d\sigma(e^+e^-)\over d\Omega dz_h}\sim
\sum_a e_a^2\,D_1^a(z_h)
\end{equation} 
with $z_h=2P_h\cdot q/Q^2$. A formal definition\cite{SoperCollins,Jaffe83} 
of the FF $D_1$ is obtained from the quark-quark correlation function
\begin{equation} 
\Delta_{ij}(k,P_h,S_h)=\sumint
\int\!\!\frac{d^{\,4}x}{(2\pi)^4}\,
e^{ik\cdot x}\,
\langle 0|\psi_i(x)|P_h,S_h;X\rangle
\langle P_h,S_h;X|\overline\psi_j(0)|0\rangle
\end{equation} 
again as (partially integrated) Dirac projection
\begin{equation} 
D_1(z) = \left.\frac{1}{4z}\int dk^+\,d^{\,2}\vec k_\sT\;
{\rm Tr}(\Delta\gamma^-) \right|_{k^- = P_h^-/z} \;.
\end{equation} 
\end{itemize} \vspace*{-4.6mm}

%%%%%%%%%%%%%%%%%%%%%%%%%%%%%%%%%%%%
\subsection{DF: partons in a hadron}
With the definition 
\begin{equation} 
\Phi^{[\Gamma]}(x) =
\left.{1\over 2}\int dp^-d^{\,2}\vec p_\sT\;
{\rm Tr}\left(\Phi\;\Gamma\right)\right|_{p^+=xP^+}
\end{equation} 
we can list all ($x$ dependent)
\ul{leading twist distribution functions} as
\begin{eqnarray}
\Phi^{[\gamma^+]}(x)                    &=& f_1(x)
\qquad \mbox{momentum distribution}\quad [\mbox{also called: }q(x)]\\
\Phi^{[\gamma^+\gamma_5]}(x)            &=& \lambda\;g_1(x)
\qquad \mbox{helicity distribution}\quad [\Delta q(x)]\\
\Phi^{[i\sigma^{\alpha +}\gamma_5]} (x) &=& S_\sT^\alpha\;h_1(x)
\qquad \mbox{transv.~spin distr.\cite{RalstonSoper}} \quad 
[\delta q(x), \Delta_\sT q(x)]
\end{eqnarray}
with $\lambda$ being the helicity and $S_\sT^\alpha$ the transverse component 
of the spin vector. Within the framework of lightcone 
quantization\cite{KogutSoperJaffe} the leading
twist distribution functions acquire a simple probabilistic 
interpretation:\\[-1.2mm]
\begin{center}
\begin{tabular}{|c|c|c|}
\hline
\multicolumn{2}{|c|}{interpretation:} & quark helicity content: \\
\hline
\multicolumn{2}{|c|}{\includegraphics[width=1.6cm]{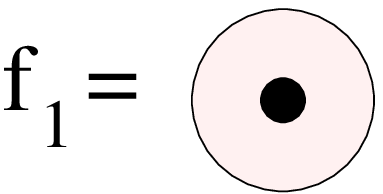}}& 
\begin{minipage}[b]{60mm} 
\begin{eqnarray*} 
\ovl{\psi}\,\g^+\,\psi &=& 
\sqrt{2}\,\psi_+^\dg \left(P_R P_R + P_L P_L \right)\;\psi_+ \\
&=& \ovl{R} R + \ovl{L} L 
\end{eqnarray*}
\end{minipage} \\
\hline
\multicolumn{2}{|c|}{\includegraphics[width=4cm]{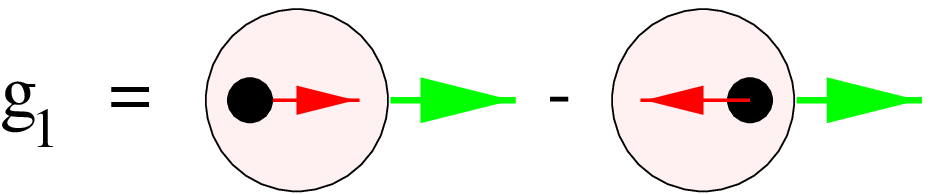}}&
\begin{minipage}[b]{60mm} 
\begin{eqnarray*} 
\ovl{\psi}\,\g^+ \g_5\,\psi &=&
 \sqrt{2}\, \psi_+^\dg \left( P_R P_R - P_L P_L \right) \psi_+ \\
&=& \ovl{R} R - \ovl{L} L 
\end{eqnarray*}
\end{minipage} \\
\hline
\includegraphics[width=2.9cm]{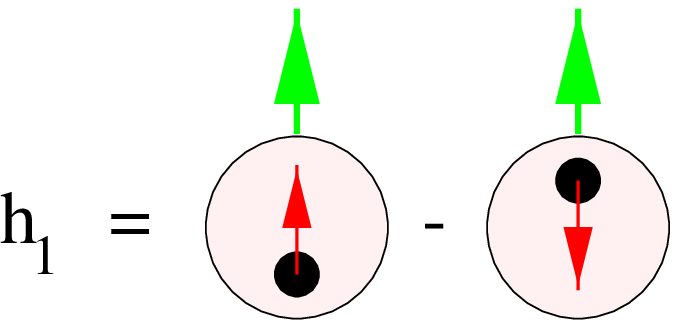}&
\multicolumn{2}{|c|}{
\begin{minipage}[b]{70mm} 
\begin{eqnarray*} 
\ovl{\psi}\,{\rm i}\sig^{i +}\g_5\,\psi
&=&\sqrt{2}\,\psi_+^{\dg}\; 
   \left( P_L \g^i P_R - P_R \g^i P_L \right) \psi_+\\
&=& \ovl{L} R - \ovl{R} L
\end{eqnarray*}
\end{minipage}} \\
\hline
\end{tabular}
\end{center} 
The DF $f_1(x)$ and $g_1(x)$ are known rather well from experiment. The 
chiral odd DF $h_1(x)$ presently is completely unknown. An obstacle to its
determination is the fact that it needs another chiral odd function to be 
combined with in an observable.

%%%%%%%%%%%%%%%%%%%%%%%%%%%%%%%%%%%%
\subsection{FF: hadrons in a parton}
Information on the hadronic structure which is complementary to the one
encoded in the distribution functions is contained in 
{\em fragmentation functions}, which describe the hadronization process
of a (current) quark to hadrons.
With the general definition
\begin{equation} 
\Delta^{[\Gamma]}(z) = \left.\frac{1}{4z}\int dk^+\,d^{\,2}\vec k_\sT\;
{\rm Tr}(\Delta\;\Gamma) \right|_{k^- = P_h^-/z} 
\end{equation} 
we can list all ($z$ dependent) \ul{leading twist fragmentation functions} as
\begin{eqnarray}
\Delta^{[\gamma^-]}(z)              &=& D_1(z)\\
\Delta^{[\gamma^-\gamma_5]}(z)      &=& \lambda_h\;G_1(z)\\
\Delta^{[i\sigma^{\alpha -}\gamma_5]} (z) 
&=& S_{h\sT}^\alpha\;H_1(z)
\end{eqnarray} 
\vspace{1mm}

\noindent
\begin{minipage}[c]{0.3\textwidth}
which have a probabilistic interpretation in analogy to the one of DF.
The only FF known from experiment is $D_1(x)$ for some species of 
hadrons, like protons, neutrons, pions and kaons.
\end{minipage} 
\begin{minipage}[c]{0.7\textwidth} 
\begin{center} 
\begin{tabular}{|c|}
\hline
interpretation:\\
\hline
\includegraphics[width=3.8cm]{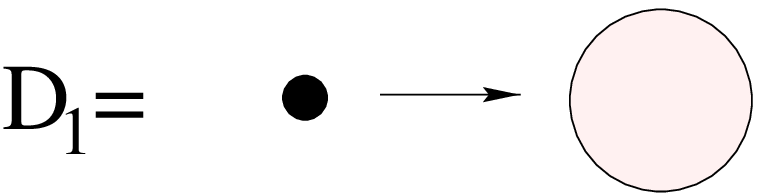} \rule[-1mm]{0mm}{11.6mm}\\
\hline
\includegraphics[width=7.2cm]{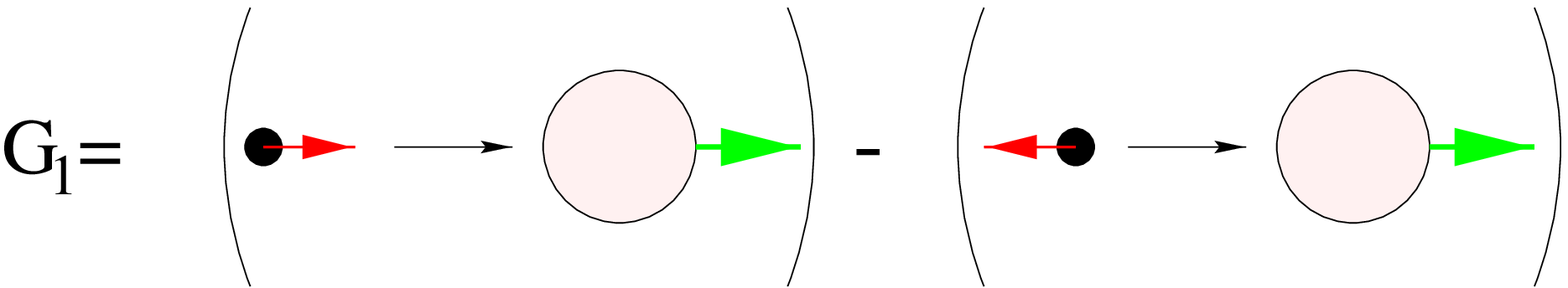} \rule[-1mm]{0mm}{15mm}\\
\hline
\includegraphics[width=6.2cm]{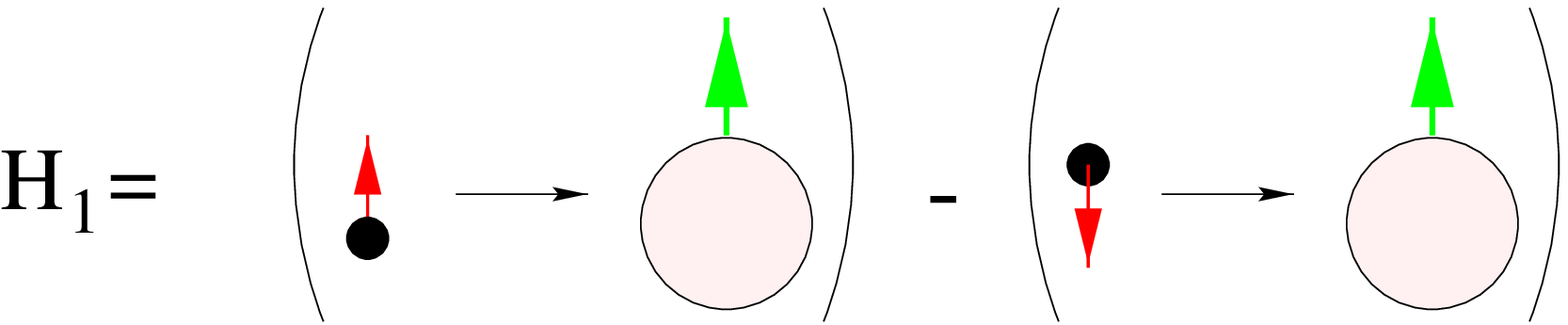} \rule[-1mm]{0mm}{15mm}\\
\hline
\end{tabular}
\end{center}
\end{minipage} 

%%%%%%%%%%%%%%%%%%%%%%%%%%%%%%%%%%%%%%%%%%%%%
\subsection{Transverse momentum dependent FF}

There are hard processes in which information on the transverse momenta
of quarks relative to their parent hadrons is retained if the observables 
are kept differential in the
transverse momentum of one of the external momenta. This leads to a larger 
number of independent DF and FF.\cite{MuldersTangerman}

Consider, for instance, the semi-inclusive DIS process: 
$\ell+H\to\ell'+h+X$, where 
one of the hadrons in the final state is observed. The leading quark 
diagram for this process is
\vspace{-1mm}
\begin{center}
\parbox[b]{40mm}{
\includegraphics[width=4cm]{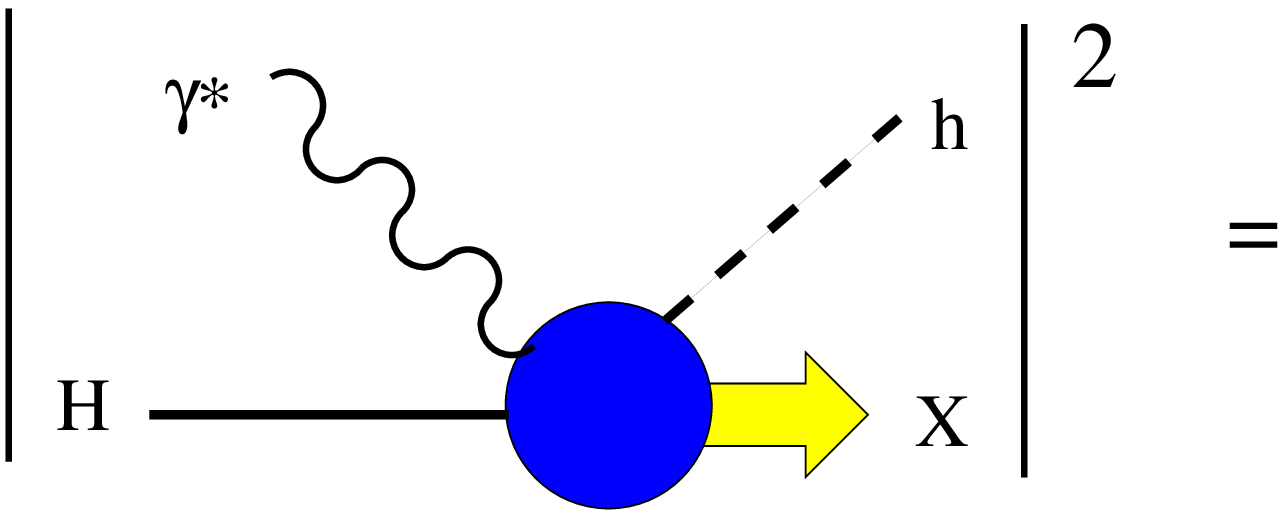}\\[6mm]}
\qquad
\includegraphics[width=4cm]{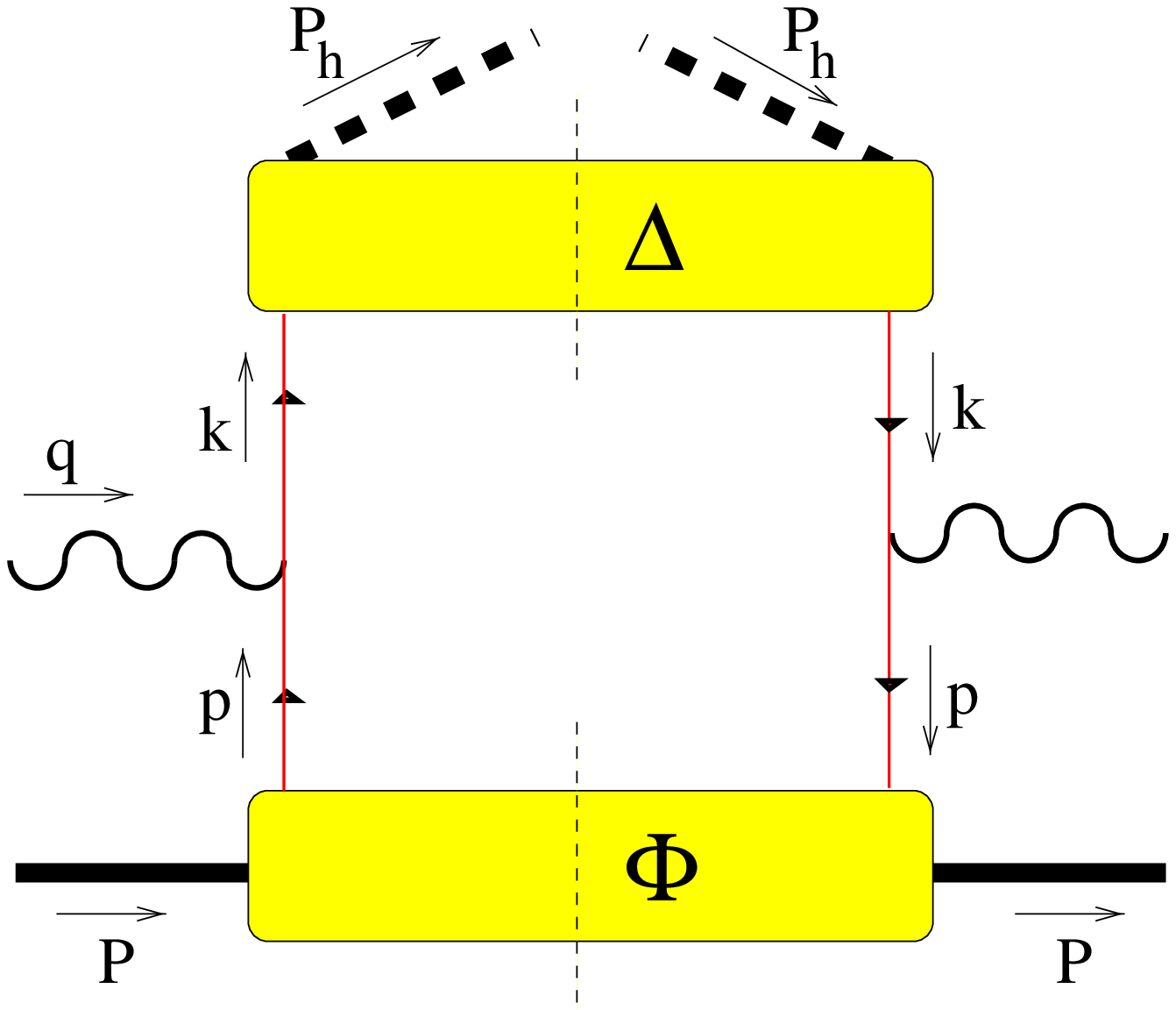}\\[-6mm]
\end{center}
The three external momenta in general are \ul{not} collinear; at least one 
of them
will have a transverse component. A possible parametrization in a frame where
target momentum and momentum of the produced hadron are collinear is
(in lightcone coordinates $a^\mu=[a^-,a^+,\vec a_\sT]$)
\[
P=\left[\frac{\xb M^2}{Q\sqrt{2}}\;,\;
        \frac{Q}{\xb \sqrt{2}}\;,\;\vec 0_\sT\right],
\quad
P_h=\left[\frac{z_hQ}{\sqrt{2}}\;,\;
          \frac{M_h^2}{z_hQ\sqrt{2}}\;,\;\vec 0_\sT\right] \;.
\]
TIn this frame the photon momentum has a transverse component
\[
q=\left[\frac{Q}{\sqrt{2}}\;,\;-\frac{Q}{\sqrt{2}}
        \;,\;\vec q_\sT\right] \;.
\]

\noindent
\begin{minipage}[b]{0.5\textwidth} 
\includegraphics[width=4cm]{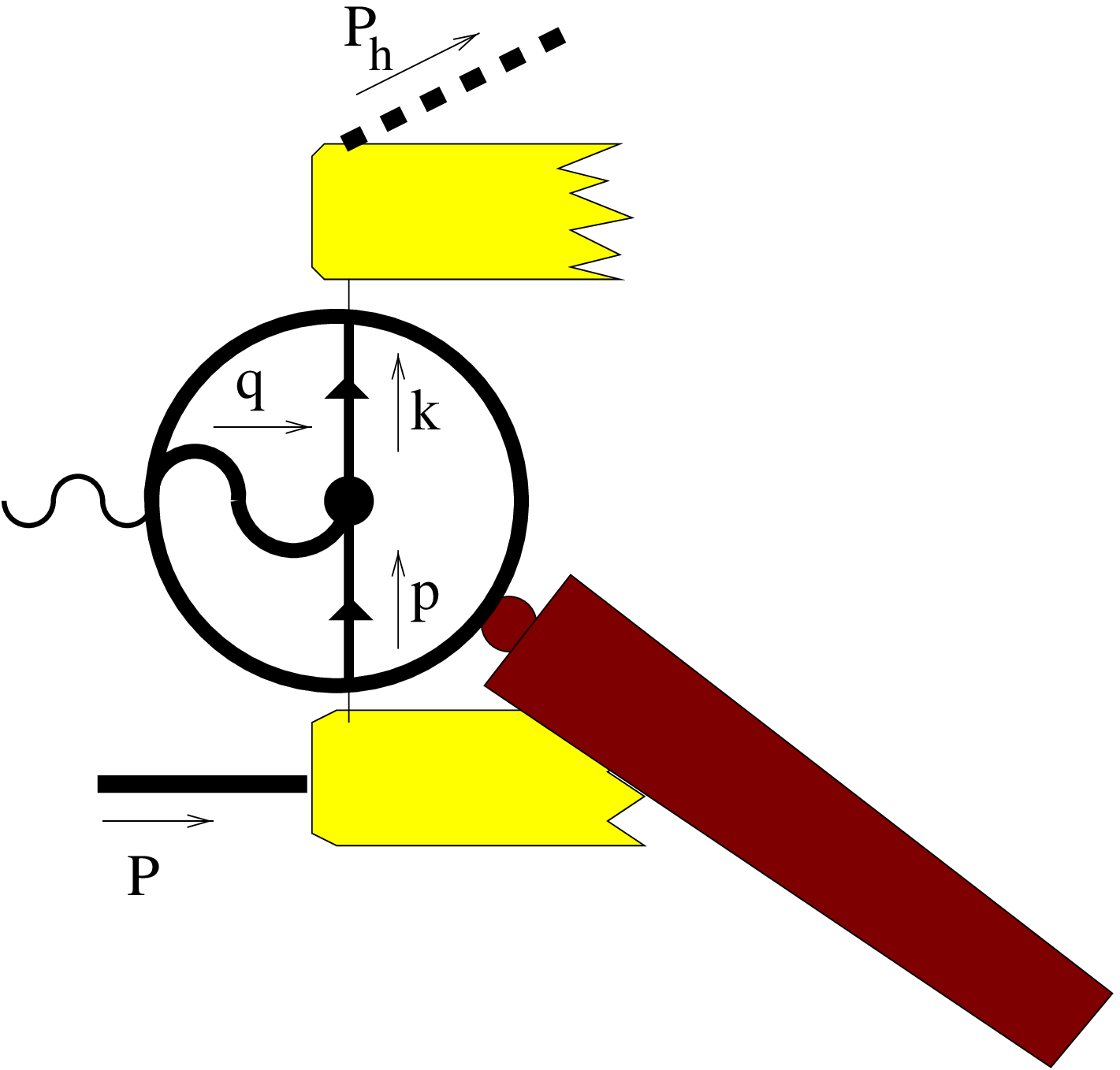}
\unitlength 1mm
\begin{picture}(0,0) 
\put(-15,25){\fbox{$\delta^2\left(\vec p_\sT+\vec q_\sT-\vec k_\sT\right)$}}
\end{picture} 
\end{minipage} 
\begin{minipage}[b]{0.5\textwidth} 
The transverse component of the photon momentum $\vec q_\sT$ is connected
to the quark transverse momenta by the momentum conservation of an elementary
vertex in the hard part of the quark diagram. Thus, the cross section
differential in the photon transverse momentum
$d\sigma/(d\ldots d^{\,2}\vec q_\sT)$ is sensitive to the quark
transverse momenta ${\vec p_\sT}$ and ${\vec k_\sT}$ relative to their
parent hadrons.
\end{minipage} 
\vspace{0.6mm}

\noindent
Quark transverse momenta relative to hadrons in DF 
and FF, respectively:\\[0.6mm]
\begin{center} 
\fbox{
\includegraphics[width=4cm]{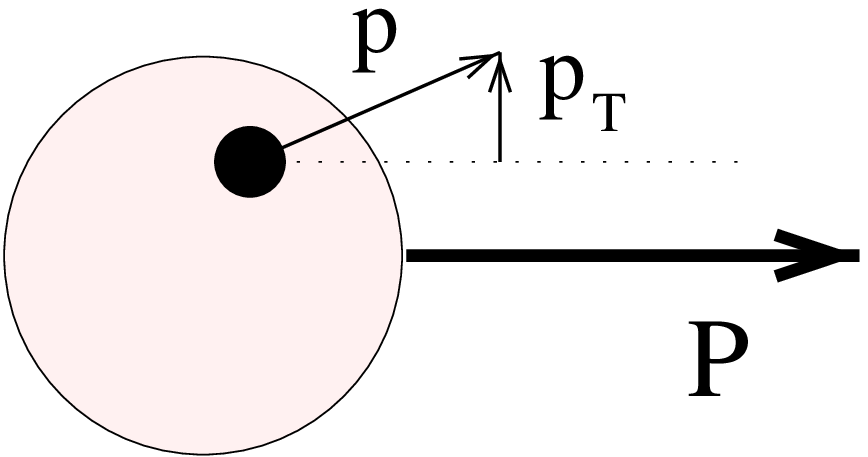}}\qquad
\fbox{
\includegraphics[width=6cm]{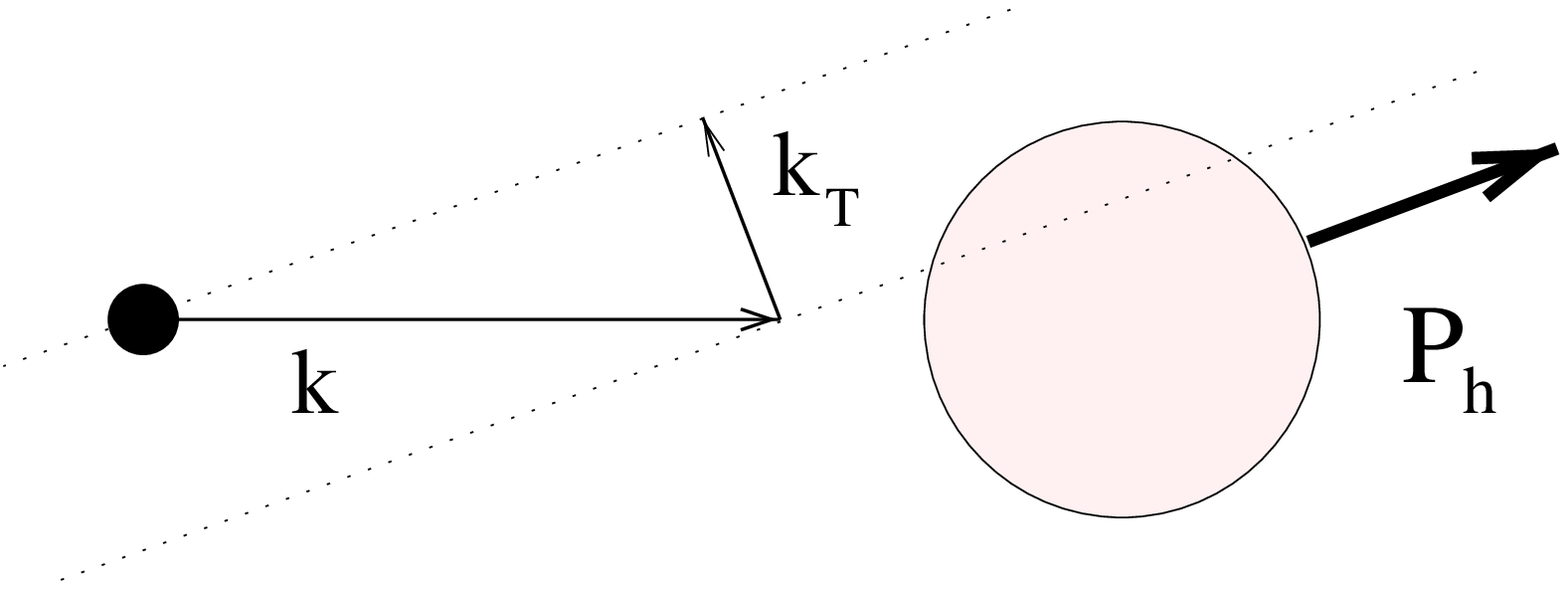}}
\end{center}
\vspace{2mm}
The leading order ($z_h$ and $\vec k_\sT$ dependent) FF are:
\begin{equation} 
\Delta^{\left[\gamma^-\right]} (z_h, \vec k_\sT) 
 = D_1 (z_h, \vec k_\sT^{\,2})
 {}+ \frac{\epsilon_{\sT ij} k^i_\sT S^j_{h\sT}}{M_h} 
     D_{1T}^{\perp} (z_h,\vec k_\sT^{\,2})
\end{equation} 
\vspace*{-5mm}
\begin{center} 
\begin{tabular}{|c|c|}
\hline
\includegraphics[width=3cm]{D1.eps}&
\includegraphics[width=3.6cm]{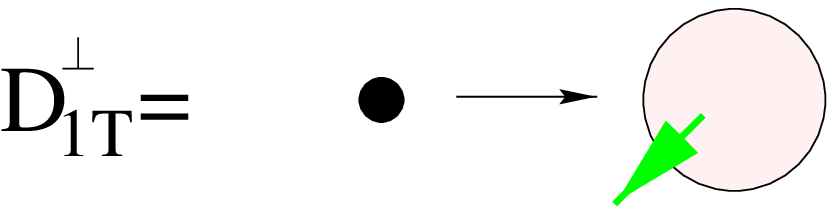}\rule{0mm}{11mm}\\[-3mm]
\rule{55mm}{0mm}&\rule{55mm}{0mm}\\
\hline
\end{tabular}
\end{center} 

\begin{equation}
\Delta^{\left[\gamma^-\gamma_5\right]} (z_h, \vec k_\sT) 
 = \lambda_h G_{1L} (z_h, \vec k_\sT^{\,2})
{}+\frac{\vec k_\sT \cdot \vec S_{h\sT}}{M_h} 
G_{1T} (z_h, \vec k_\sT^{\,2})
\end{equation} 
\vspace*{-5mm}
\begin{center} 
\begin{tabular}{|c|c|}
\hline
\includegraphics[width=55mm]{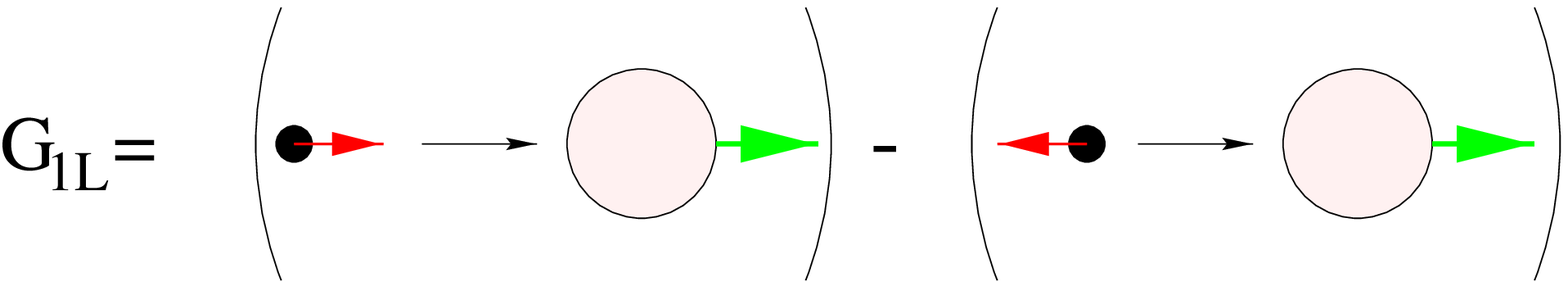}&
\includegraphics[width=55mm]{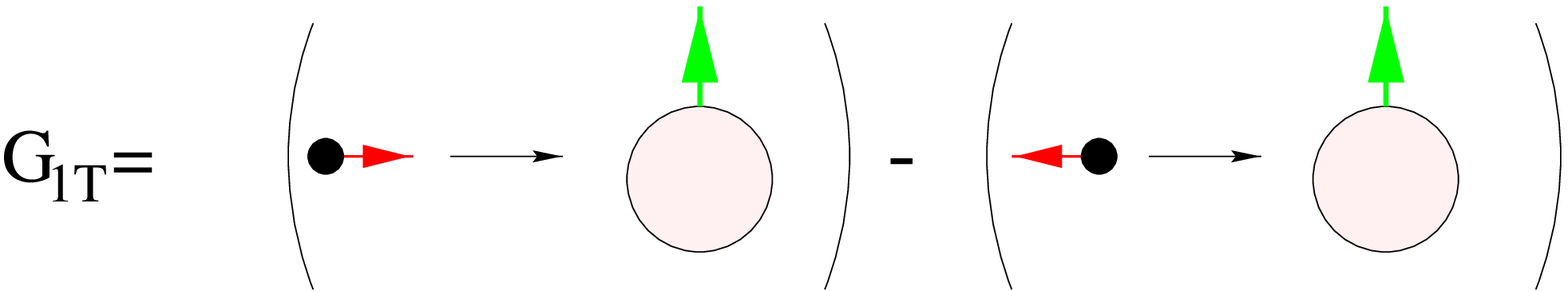}\rule[-1mm]{0mm}{12.4mm}\\
\hline
\end{tabular}
\end{center} 

\begin{eqnarray}
\lefteqn{ 
\Delta^{\left[{\rm i} \sigma^{i -} \gamma_5\right]} 
(z_h, \vec k_\sT) = S_{h\sT}^i H_1 (z_h, \vec k_\sT^{\,2})
{}+\frac{\epsilon_\sT^{i j} k_{\sT j}}{M_h} 
H_1^{\perp} (z_h, \vec k_\sT^{\,2})} \nn\\[2mm]
&&{}+\frac{\lambda_h k_\sT^i}{M_h} 
H_{1L}^{\perp} (z_h, \vec k_\sT^{\,2})
  {}+\frac{\left( k_\sT^i k_\sT^j 
   - \vec k_\sT^{\,2} \delta_{ij}/2 \right) S_{h\sT}^j}{M_h^2} 
H_{1T}^{\perp} (z_h, \vec k_\sT^{\,2})
\end{eqnarray} 
\vspace*{-5mm}
\begin{center} 
\begin{tabular}{|c|c|}
\hline
\includegraphics[width=55mm]{H1.eps}&
\includegraphics[width=55mm]{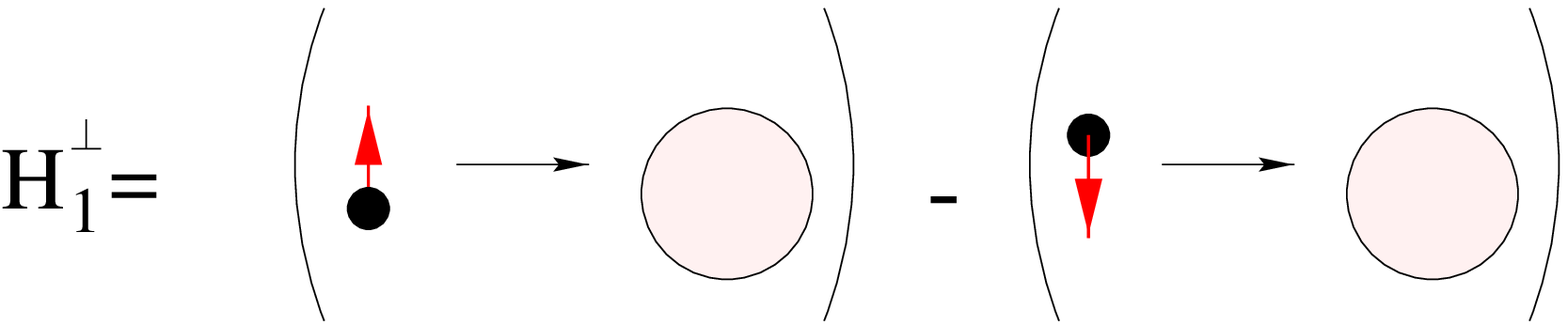}\rule[-1mm]{0mm}{12.4mm}\\
\hline
\includegraphics[width=55mm]{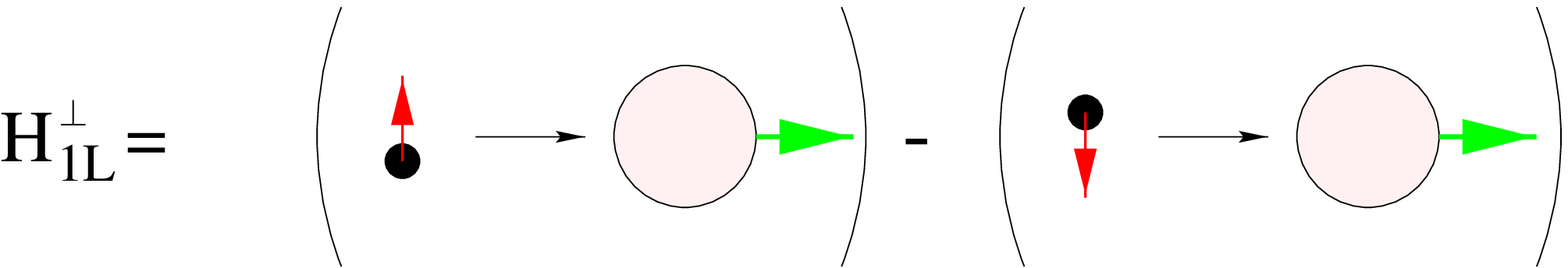}&
\includegraphics[width=55mm]{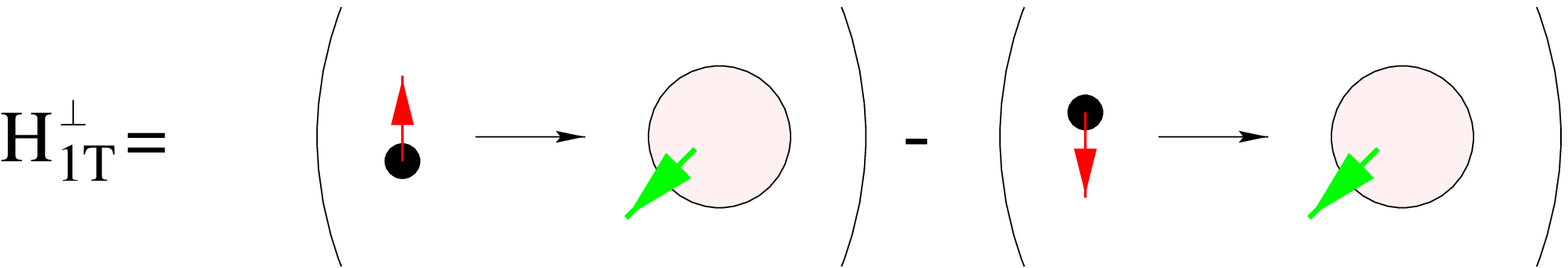}\rule[-1mm]{0mm}{12.4mm}\\
\hline
\end{tabular}
\end{center}

\vspace{2mm}
The FF $D_{1T}^\perp$ and $H_1^\perp$ are time reversal odd (T-odd). Here 
the notion
``time reversal odd'' is used in the sense: {\em ``in the absence of
final state interactions (FSI) those functions would be
forbidden by a constraint from time reversal invariance''}.

The interest in T-odd fragmentation functions was triggered 
by a proposal to measure the transversity distribution $h_1$ from an 
azimuthal asymmetry in one-hadron inclusive DIS which involves 
$H_1^\perp$ as the necessary chiral odd partner~\cite{col93}
\begin{equation} 
\frac{d \sigma}{d \ldots d^2 \vec q_\sT} \sim \  
\sin (\phi) \  h_1 (x, \vec p_\sT^{\,\,2}) \  
H_1^{\perp} (z_h, \vec k_\sT^{\,2}) \;.
\end{equation} 

\section{Two-hadron fragmentation functions}
\subsection{Why two-hadron FF ?}
The modeling of T-odd one-hadron FF with specific assumptions on the
hadronizatin process encounters serious difficulties. A necessary ingredient
is the existence of at least two competing channels interfering through a
non-vanishing phase. Moreover, it turns out that a genuine difference in 
the Lorentz structure of the vertices describing the fragmentation is needed.

On general grounds, model assumptions for an interaction of the observed
hadron with the rest of the jet have either problems with factorization
breaking, with translational and rotational invariance, or they can be
redefined in the quark/hadron/rest-of-jet vertex, thus not resulting in the
necessary modification of the Lorentz structure. Moreover, it was also 
argued\cite{jafjitang} that the required sum over all possible states of 
the fragments could average out the FSI effects.

These arguments naturally lead to the consideration of final state 
interactions between two hadrons emitted inside the same jet as a source for
T-odd fragmentation functions. 

For the case of the two hadrons being a pair of pions the resulting FF 
have been proposed to investigate the transverse spin dependence of 
fragmentation. Collins and Ladinsky\cite{collad} considered the 
interference of a scalar resonance with the channel of independent 
successive two pion production. Jaffe, Jin and Tang\cite{jafjitang} 
proposed the interference of $s$- and $p$-wave production channels, where 
the relevant phase shifts are essentially known.

\subsection{Definition of two-hadron FF and symmetry properties}
Consider the situation where a pair of hadrons is produced in the same 
quark current jet: $l H \rightarrow l' h_1 h_2 X$. By generalizing the 
Collins-Soper light-cone formalism\cite{SoperCollins} for fragmentation into 
multiple hadrons the corresponding quark-quark correlation function 
can be defined
\begin{equation} 
\Delta(k;P_1,P_2)=
\sumint \int\!\!\frac{d^{\,4\!}\zeta}{(2\pi)^4}\;
e^{ik\cdot\zeta}\;
\langle 0|\psi_i(\zeta)|h_1,h_2,X\rangle
\langle X,h_1,h_2|\ovl{\psi}_j(0)|0\rangle
\end{equation} 

\begin{minipage}[c]{35mm} 
\begin{center}
\epsfig{file=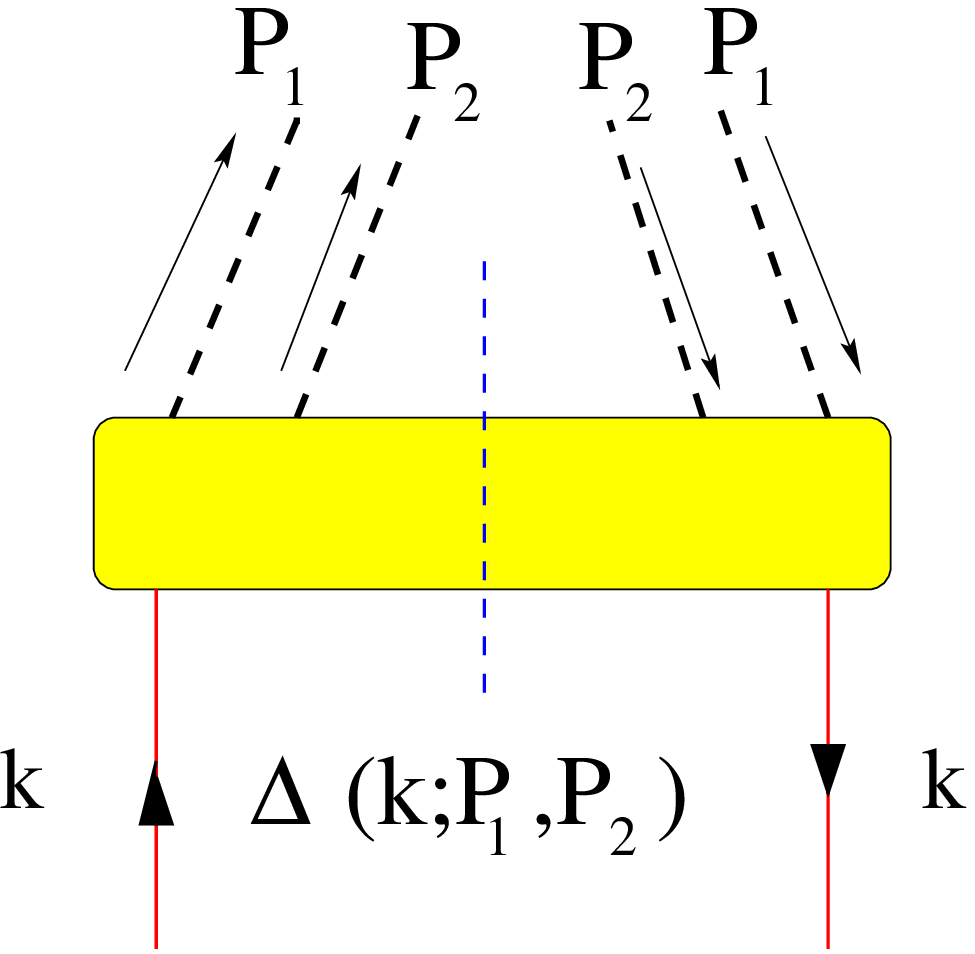,width=3.4cm}
\end{center}
\end{minipage}\hspace{5mm}
\begin{minipage}[c]{70mm} 
For the corresponding
quark-quark correlation function $\Delta(k;P_1,P_2)$ one can make the most
general ansatz (allowed by {\em parity invariance})
\begin{eqnarray}
\lefteqn{
\Delta(k;P_1,P_2)=}\nn\\[2mm]
&&
C_1\,\left(M_1+M_2\right) 
+ C_2\,\Pslash_1  
+ C_3\,\Pslash_2
+ C_4\,\kslash\nn\\
&&
{}+ \frac{C_{5}}{M_1}\,\sig^{\mu \nu} P_{1\mu} k_\nu
{}+ \frac{C_{6}}{M_2}\,\sig^{\mu \nu} P_{2\mu} k_\nu \nn\\
&&
{}+ \frac{C_{7}}{M_1+M_2}\,\sig^{\mu \nu} P_{1\mu} P_{2\nu}\nn\\
&&
{}+ \frac{C_8}{M_1 M_2}\,\g_5\,\eps^{\mu\nu\rho\sig}
     \g_\mu P_{1\nu} P_{2\rho} k_\sig \;.
\label{eq:ansatz}
\end{eqnarray} 
\end{minipage}\\[4mm]
From {\em hermiticity} follows $C_i^* = C_i$ for $i=1-4,5-8$, and 
a constraint from {\em time-reversal invariance} (if applicable) leads to
$C^*_i = C_i$ for $i=1-4$ and $C^*_i = - C_i$ for $i=5-8$. FSI, however, 
render the constraint inapplicable, which otherwise would require 
$C_5..C_8=0$. FF involving $C_5..C_8$ are called T-odd.

Before defining the two-hadron FF 
some kinematic quantities have to be introduced:\\
\noindent
\begin{minipage}[t]{55mm} 
\begin{itemize}
\item
sum of four-momenta of the hadron pair: $P_h=P_1+P_2$
\end{itemize} 
\end{minipage}  \qquad 
\begin{minipage}[t]{50mm} 
\begin{itemize} 
\item
momentum fractions:\\
$ P_1^- = \xi\;z_h\;k^- $ \\ $P_2^- = (1-\xi)\;z_h\;k^-$
\end{itemize} 
\end{minipage}\\[-8mm] 
\begin{itemize} 
\item
\begin{minipage}[t]{45mm}
(half of) relative transverse momentum inside the hadron pair:\\
$\vec R_\sT=(\vec P_{1\sT}-\vec P_{2\sT})/2$
\end{minipage} \quad
\begin{minipage}{60mm} 
\begin{center} 
\raisebox{-20mm}[20mm]{\includegraphics[width=55mm]{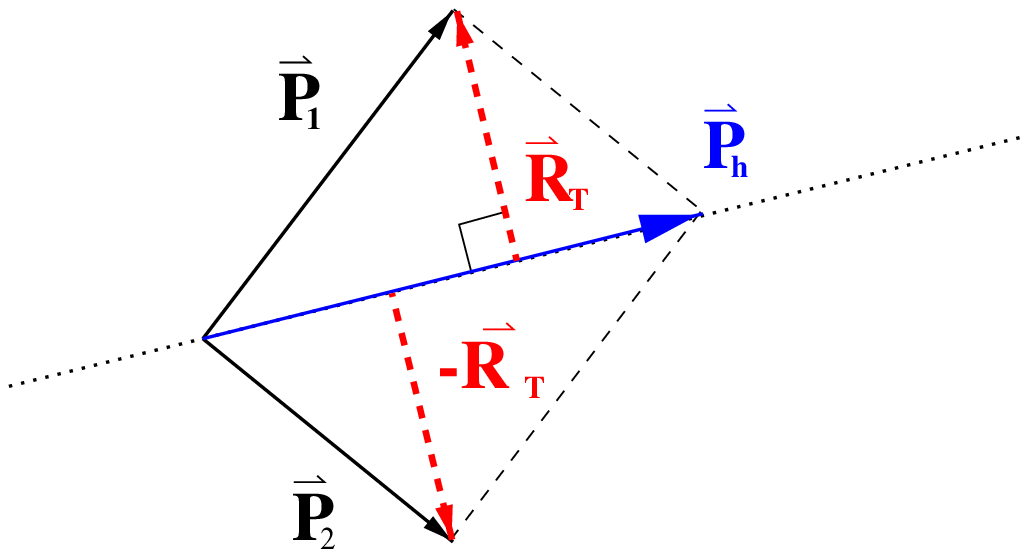}}
\end{center}
\end{minipage}
\item
relative transverse momentum between quark and hadron pair: $\vec k_T$\\
\begin{center}
\includegraphics[width=75mm]{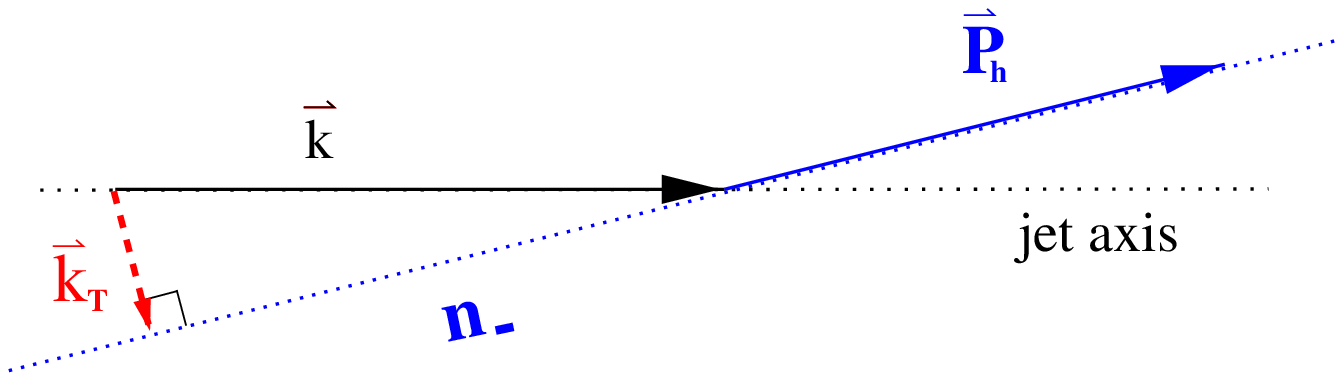}
\end{center}
\end{itemize} 
The two-hadron FF are defined as projections of $\Delta(k;P_1,P_2)$
on different Dirac structures. They depend on five variables: on the momentum
fraction of the hadron pair $(z_h)$, on the way this momentum is shared 
inside the pair $(\xi)$, on the invariant mass of the pair $M_h^2$, and on 
the ``geometry'' of the pair, namely on the relative orientation between 
the hadron pair plane and the quark jet axis 
(${\vec k}_\sT^{\,2}$, ${\vec k}_\sT \cdot {\vec R}_\sT$).\\[3mm]

\noindent
To leading order there are\\
\begin{minipage}{0.6\textwidth} 
\begin{equation} 
\Delta^{[\gamma^-]}
(z_h,\xi,{\vec k}_\sT^{\,2},\vec k_\sT\cdot\vec R_\sT,M_h^2) =  D_1
\end{equation} 
\end{minipage} 
\begin{minipage}{0.39\textwidth}
\begin{center}
\fbox{\includegraphics[width=34mm]{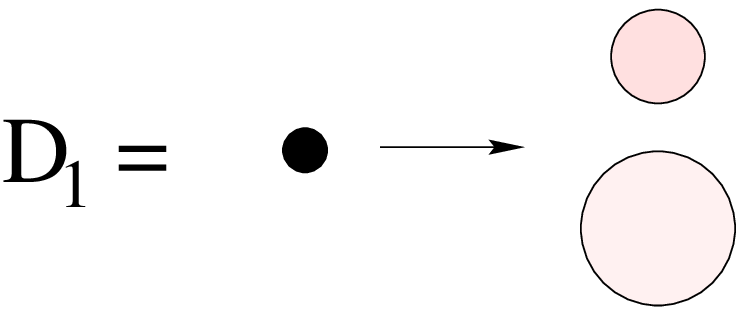}}
\end{center} 
\end{minipage}\\[2mm] 
and the interference FF
\begin{equation} 
\Delta^{\left[\gamma^- \gamma_5\right]}
(z_h,\xi,{\vec k}_\sT^{\,2},\vec k_\sT\cdot\vec R_\sT,M_h^2) = 
\frac{\epsilon_\sT^{ij}\,R_{\sT i}\,k_{\sT j}}{M_p M_{\pi}}\, G_1^{\perp}
\end{equation} 
\begin{center}
\fbox{\parbox{0.98\textwidth}{
\begin{center} \includegraphics[width=70mm]{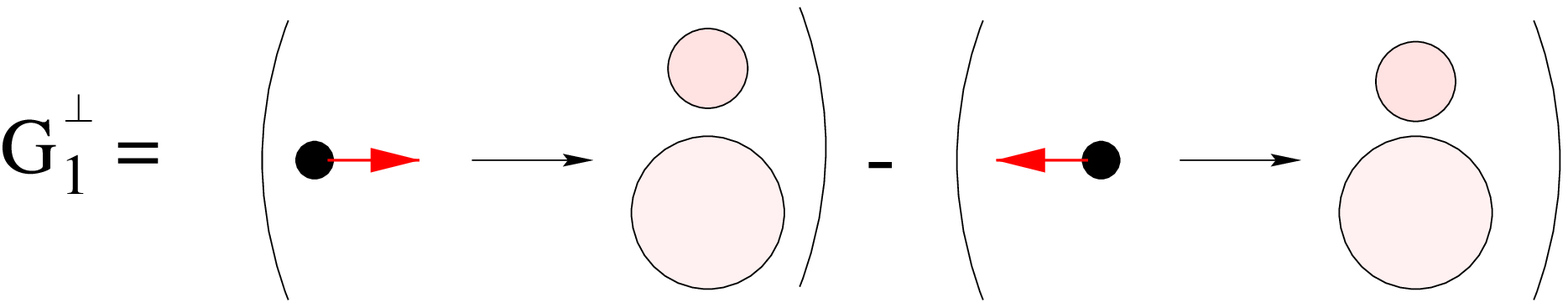}
\end{center} }}
\end{center}
\vspace*{2mm}
and
\begin{equation} 
\Delta^{\left[{\rm i} \sigma^{i-} \gamma_5\right]} 
(z_h,\xi,{\vec k}_\sT^{\,2},\vec k_\sT\cdot\vec R_\sT,M_h^2) =
\frac{\epsilon_\sT^{ij}\,R_{\sT j}}{M_p + M_{\pi}}\,H_1^{\newangle}\; +
\frac{\epsilon_\sT^{ij}\,k_{\sT j}}{M_p + M_{\pi}}\,H_1^{\perp}
\end{equation} 
\begin{center}
\fbox{\parbox{0.98\textwidth}{
\begin{center} \includegraphics[width=70mm]{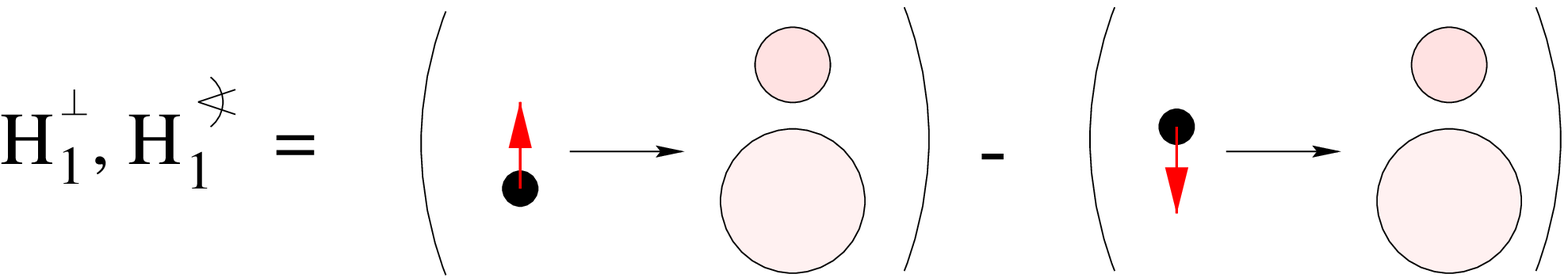}
\end{center} }}
\end{center}

\vspace*{3mm}
Inserting the ansatz (\ref{eq:ansatz}) for $\Delta(k;P_1,P_2)$ reveals that
$G_1^{\perp}$, $H_1^{\newangle}$ and $H_1^{\perp}$ are T-odd. From the
Dirac matrices in the projections it can be deduced that $D_1$ and 
$G_1^{\perp}$ are chiral even, and $H_1^{\newangle}$, $H_1^{\perp}$ are 
chiral odd.

%%%%%%%%%%%%%%%%%%%%%%%%%%%%%%%%%%%%%%%%
\subsection{Pion-Proton interference FF}
Recently, we have calculated explicitely the two-hadron interference 
fragmentation functions for the case of the hadron pair being a pion and a 
proton (produced in the same jet)\cite{us}. The calculation is done in an 
extended version of the spectator model used in Ref.~\cite{spectator}.
The interference of two channels is considered, where the hadron pair
is produced either through a Roper resonance, or as successive independent 
production.

The basic idea of the spectator model is to make a specific ansatz for the
spectral decomposition of the quark correlator by replacing the sum over 
the complete set of intermediate states in (\ref{eq:ansatz}) with an 
effective spectator state with a definite mass $M_D$ and the quantum 
numbers of the diquark.

As a consequence of the quark-quark correlation function simplifies to a form
\begin{eqnarray} 
\Delta_{ij}(k;P_p,P_\pi)&=&
\frac{\theta\!\left((k-P_h)^+\right)}{(2\pi)^3} \; 
\delta\left((k-P_h)^2-M_D^2\right)\nn\\
&&\hspace{12mm}\times
\langle 0|\psi_i(0)|\pi,p,D\rangle\langle D,p,\pi|\ovl{\psi}_j(0)|0\rangle 
\end{eqnarray} 
containing amplitudes which can be calculated from Feynman 
diagrams. There are diagonal diagrams contributing to the quark 
fragmentation
\begin{center}
\unitlength=0.6mm
%
% diag Ia
%
\begin{picture}(60,45)
\put(10,15){\circle*{2}} % qDR vert
\put(5,5){\line(1,2){5}}\put(5,5){\vector(1,2){3}} % q prop
\multiput(10,14.5)(0,1){2}{\line(1,0){20}} % D prop
\put(10,29){\circle*{2}} % Rppi vert
\put(2,40){$p$}\put(18,40){$\pi$} % text
\put(1,8){$q$}\put(3,22){$R$}\put(18,9){$D$} % text
 \multiput(10.0,29.0)(1.4,1.4){7}{\circle*{0.5}} % pi prop
 \multiput(10.2,29.2)(1.4,1.4){7}{\circle*{0.5}} % pi prop
 \multiput(10.4,29.4)(1.4,1.4){7}{\circle*{0.5}} % pi prop
\thicklines
 \put(10,29){\line(-1,1){8}}\put(10,29){\vector(-1,1){5}} % p prop
% \multiput(10,29)(1.4,1.4){6}{\line(1,1){0.7}} % pi prop
 \put(9.5,15){\line(0,1){14}}              % Roper prop
 \multiput(10.5,15)(0,2){8}{\line(0,1){1}} % Roper prop
\thinlines
%%%
\put(30,4){\line(0,1){30}}
\put(27,4){\oval(6,6)[br]}%{\arc{6}{0}{1.57}}
\put(33,34){\oval(6,6)[tl]}%{\arc{6}{3.14}{4.71}}
%%%
\put(50,15){\circle*{2}} % qDR vert
\put(50,15){\line(1,-2){5}}\put(50,15){\vector(1,-2){3}} % q prop
\multiput(30,14.5)(0,1){2}{\line(1,0){20}} % D prop
\put(50,29){\circle*{2}} % Rppi vert
\put(57,40){$p$}\put(41,40){$\pi$} % text
\put(57,8){$q$}\put(53,22){$R$}\put(39,9){$D$} % text
 \multiput(50.0,29.0)(-1.4,1.4){7}{\circle*{0.5}} % pi prop
 \multiput(49.8,29.2)(-1.4,1.4){7}{\circle*{0.5}} % pi prop
 \multiput(49.6,29.4)(-1.4,1.4){7}{\circle*{0.5}} % pi prop
\thicklines
 \put(50,29){\line(1,1){8}}\put(58,37){\vector(-1,-1){5}} % p prop
% \multiput(50,29)(-1.4,1.4){6}{\line(-1,1){0.7}} % pi prop
 \put(50.5,15){\line(0,1){14}}              % Roper prop
 \multiput(49.5,15)(0,2){8}{\line(0,1){1}} % Roper prop
\thinlines
\end{picture} \quad
%
% diag Ib
%
\begin{picture}(60,45)
\put(10,15){\circle*{2}} % qpD vert
\put(5,5){\line(1,2){5}}\put(5,5){\vector(1,2){3}} % q prop
\put(20,15){\circle*{2}} % DpiD vert
\multiput(20,14.5)(0,1){2}{\line(1,0){10}} % D prop
\put(10,15){\line(1,0){10}} % q prop
\put(19,40){$p$}\put(9,40){$\pi$}\put(13,9){$q$} % text
\thicklines
 \multiput(10,15)(0,2){10}{\line(0,1){1}} % pi prop
 \put(20,15){\line(0,1){19}}\put(20,15){\vector(0,1){12}} % p prop
\thinlines
%%%
\put(30,4){\line(0,1){30}}
\put(27,4){\oval(6,6)[br]}%{\arc{6}{0}{1.57}}
\put(33,34){\oval(6,6)[tl]}%{\arc{6}{3.14}{4.71}}
%%%
\put(50,15){\circle*{2}} % qpD vert
\put(55,5){\line(-1,2){5}}\put(50,15){\vector(1,-2){3}} % q prop
\put(40,15){\circle*{2}} % DpiD vert
\multiput(30,14.5)(0,1){2}{\line(1,0){10}} % D prop
\put(40,15){\line(1,0){10}} % q prop
\put(39,40){$p$}\put(49,40){$\pi$}\put(43,9){$q$} % text
\thicklines
 \multiput(50,15)(0,2){10}{\line(0,1){1}} % pi prop
 \put(40,15){\line(0,1){19}}\put(40,34){\vector(0,-1){12}} % p prop
\thinlines
\end{picture} \quad
%
% diag Ic
%
\begin{picture}(60,45)
\put(10,15){\circle*{2}} % qpD vert
\put(5,5){\line(1,2){5}}\put(5,5){\vector(1,2){3}} % q prop
\put(20,15){\circle*{2}} % DpiD vert
\multiput(10,14.5)(0,1){2}{\line(1,0){20}} % D prop
\put(19,40){$\pi$}\put(9,40){$p$}\put(13,9){$D$} % text
\thicklines
 \multiput(20,15)(0,2){10}{\line(0,1){1}} % pi prop
 \put(10,15){\line(0,1){19}}\put(10,15){\vector(0,1){12}} % p prop
\thinlines
%%%
\put(30,4){\line(0,1){30}}
\put(27,4){\oval(6,6)[br]}%{\arc{6}{0}{1.57}}
\put(33,34){\oval(6,6)[tl]}%{\arc{6}{3.14}{4.71}}
%%%
\put(50,15){\circle*{2}} % qpD vert
\put(55,5){\line(-1,2){5}}\put(50,15){\vector(1,-2){3}} % q prop
\put(40,15){\circle*{2}} % DpiD vert
\multiput(30,14.5)(0,1){2}{\line(1,0){20}} % D prop
\put(39,40){$\pi$}\put(49,40){$p$}\put(43,9){$D$} % text
\thicklines
 \multiput(40,15)(0,2){10}{\line(0,1){1}} % pi prop
 \put(50,15){\line(0,1){19}}\put(50,34){\vector(0,-1){12}} % p prop
\thinlines
\end{picture}\\
$D-a$ \hspace{30mm} $D-b$ \hspace{30mm} $D-c$
\end{center}
and interference diagrams
\begin{center}
\unitlength=0.6mm
%
% diag IIa
%
\begin{picture}(60,45)
\put(10,15){\circle*{2}} % qDR vert
\put(5,5){\line(1,2){5}}\put(5,5){\vector(1,2){3}} % q prop
\multiput(10,14.5)(0,1){2}{\line(1,0){20}} % D prop
\put(10,29){\circle*{2}} % Rppi vert
\put(2,40){$p$}\put(18,40){$\pi$} % text
\put(1,8){$q$}\put(3,22){$R$}\put(18,9){$D$} % text
\thicklines
 \put(10,29){\line(-1,1){8}}\put(10,29){\vector(-1,1){5}} % p prop
% \multiput(10,29)(1.4,1.4){6}{\line(1,1){0.7}} % pi prop
 \multiput(10.0,29.0)(1.4,1.4){7}{\circle*{0.5}} % pi prop
 \multiput(10.2,29.2)(1.4,1.4){7}{\circle*{0.5}} % pi prop
 \multiput(10.4,29.4)(1.4,1.4){7}{\circle*{0.5}} % pi prop
 \put(9.5,15){\line(0,1){14}}              % Roper prop
 \multiput(10.5,15)(0,2){8}{\line(0,1){1}} % Roper prop
\thinlines
%%%
\put(30,4){\line(0,1){30}}
\put(27,4){\oval(6,6)[br]}%{\arc{6}{0}{1.57}}
\put(33,34){\oval(6,6)[tl]}%{\arc{6}{3.14}{4.71}}
%%%
\put(50,15){\circle*{2}} % qpD vert
\put(55,5){\line(-1,2){5}}\put(50,15){\vector(1,-2){3}} % q prop
\put(40,15){\circle*{2}} % DpiD vert
\multiput(30,14.5)(0,1){2}{\line(1,0){20}} % D prop
\put(39,40){$\pi$}\put(49,40){$p$}\put(43,9){$D$} % text
\thicklines
 \multiput(40,15)(0,2){10}{\line(0,1){1}} % pi prop
 \put(50,15){\line(0,1){19}}\put(50,34){\vector(0,-1){12}} % p prop
\thinlines
\end{picture} \quad
%
% diag IIb
%
\begin{picture}(60,45)
\put(10,15){\circle*{2}} % qpD vert
\put(5,5){\line(1,2){5}}\put(5,5){\vector(1,2){3}} % q prop
\put(20,15){\circle*{2}} % DpiD vert
\multiput(10,14.5)(0,1){2}{\line(1,0){20}} % D prop
\put(19,40){$\pi$}\put(9,40){$p$}\put(13,9){$D$} % text
\thicklines
 \multiput(20,15)(0,2){10}{\line(0,1){1}} % pi prop
 \put(10,15){\line(0,1){19}}\put(10,15){\vector(0,1){12}} % p prop
\thinlines
%%%
\put(30,4){\line(0,1){30}}
\put(27,4){\oval(6,6)[br]}%{\arc{6}{0}{1.57}}
\put(33,34){\oval(6,6)[tl]}%{\arc{6}{3.14}{4.71}}
%%%
\put(50,15){\circle*{2}} % qpD vert
\put(55,5){\line(-1,2){5}}\put(50,15){\vector(1,-2){3}} % q prop
\put(40,15){\circle*{2}} % DpiD vert
\multiput(30,14.5)(0,1){2}{\line(1,0){10}} % D prop
\put(40,15){\line(1,0){10}} % q prop
\put(39,40){$p$}\put(49,40){$\pi$}\put(43,9){$q$} % text
\thicklines
 \multiput(50,15)(0,2){10}{\line(0,1){1}} % pi prop
 \put(40,15){\line(0,1){19}}\put(40,34){\vector(0,-1){12}} % p prop
\thinlines
\end{picture} \quad
%
% diag IIc
%
\begin{picture}(60,45)
\put(10,15){\circle*{2}} % qDR vert
\put(5,5){\line(1,2){5}}\put(5,5){\vector(1,2){3}} % q prop
\multiput(10,14.5)(0,1){2}{\line(1,0){20}} % D prop
\put(10,29){\circle*{2}} % Rppi vert
\put(2,40){$p$}\put(18,40){$\pi$} % text
\put(1,8){$q$}\put(3,22){$R$}\put(18,9){$D$} % text
\thicklines
 \put(10,29){\line(-1,1){8}}\put(10,29){\vector(-1,1){5}} % p prop
% \multiput(10,29)(1.4,1.4){6}{\line(1,1){0.7}} % pi prop
 \multiput(10.0,29.0)(1.4,1.4){7}{\circle*{0.5}} % pi prop
 \multiput(10.2,29.2)(1.4,1.4){7}{\circle*{0.5}} % pi prop
 \multiput(10.4,29.4)(1.4,1.4){7}{\circle*{0.5}} % pi prop
 \put(9.5,15){\line(0,1){14}}              % Roper prop
 \multiput(10.5,15)(0,2){8}{\line(0,1){1}} % Roper prop
\thinlines
%%%
\put(30,4){\line(0,1){30}}
\put(27,4){\oval(6,6)[br]}%{\arc{6}{0}{1.57}}
\put(33,34){\oval(6,6)[tl]}%{\arc{6}{3.14}{4.71}}
%%%
\put(50,15){\circle*{2}} % qpD vert
\put(55,5){\line(-1,2){5}}\put(50,15){\vector(1,-2){3}} % q prop
\put(40,15){\circle*{2}} % DpiD vert
\multiput(30,14.5)(0,1){2}{\line(1,0){10}} % D prop
\put(40,15){\line(1,0){10}} % q prop
\put(39,40){$p$}\put(49,40){$\pi$}\put(43,9){$q$} % text
\thicklines
 \multiput(50,15)(0,2){10}{\line(0,1){1}} % pi prop
 \put(40,15){\line(0,1){19}}\put(40,34){\vector(0,-1){12}} % p prop
\thinlines
\end{picture}\\
$I-a$ \hspace{31mm} $I-b$ \hspace{31mm} $I-c$
\end{center} 
On the Roper resonance diagram D-a dominates the $D_1$ function,
and I-a, I-c give the dominant contributions to the interference 
functions $G_1^\perp$, $H_1^\newangle$, $H_1^\perp$.

For the explicit calculation all propagators and vertices occurring in the 
diagrams have to be specified. Particularly important are form factors at the
vertices, which prevent the propagators from having large virtualities. The
asymptotic behavior of the form factors has been deduced from quark counting 
rules, the strength of the couplings can be determined from phenomenological
considerations. Details of this model calculation will be published 
elsewhere.\cite{us}

\section{Numerical results}

For plotting the results we assume that the proton-pion pair has an invariant
mass equal to the Roper resonance $M_h = M_R = 1.44$ GeV, and a special 
kinematic situation is chosen, where $\vec k_\sT \cdot \vec R_\sT = 0$ 
holds. Then the dependences reduce 
to $G_1^\perp (z_h, \xi, \vec k_\sT^{\,2})$, etc.

\begin{center} 
\includegraphics[width=55mm]{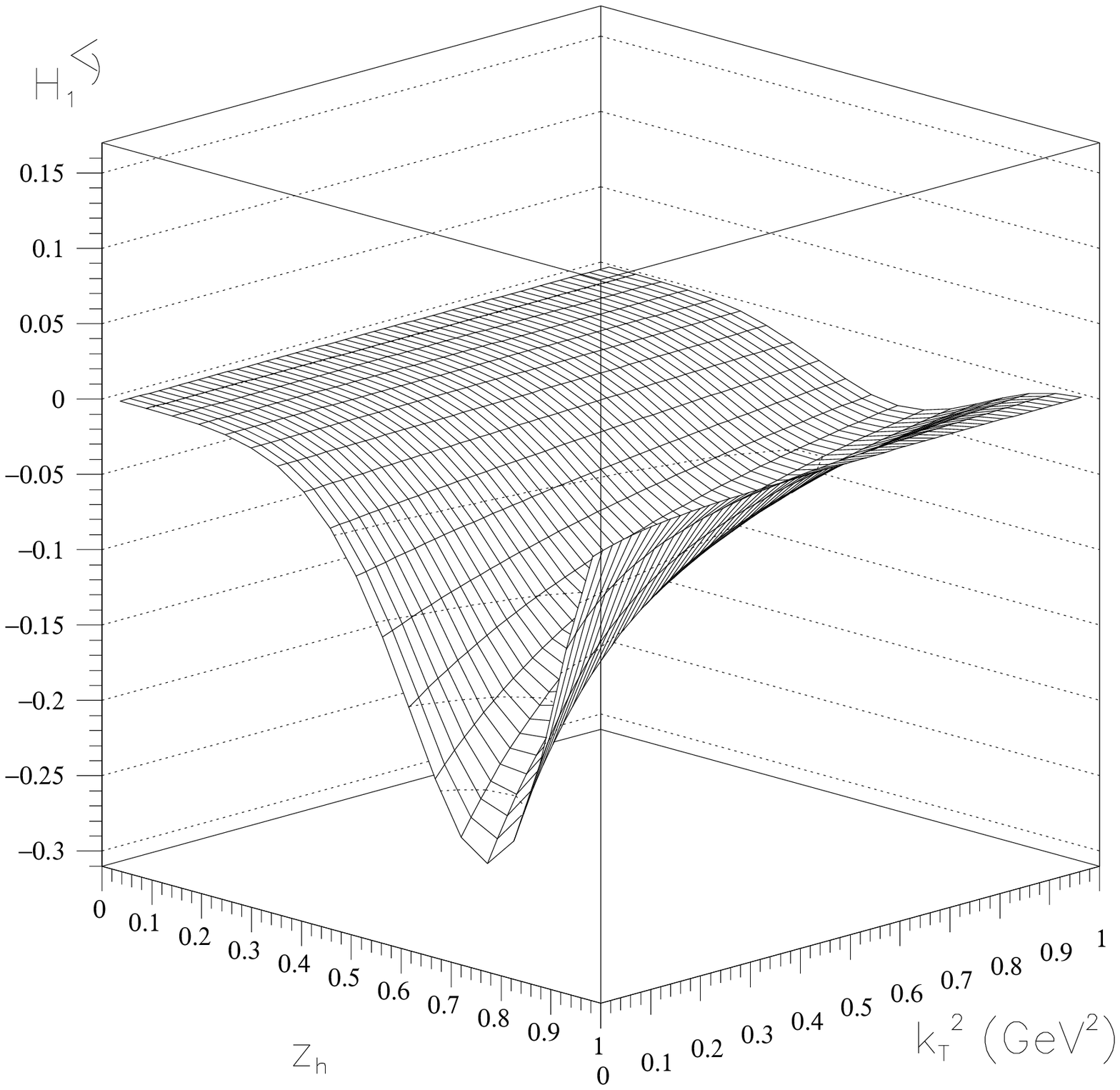} 
\includegraphics[width=55mm]{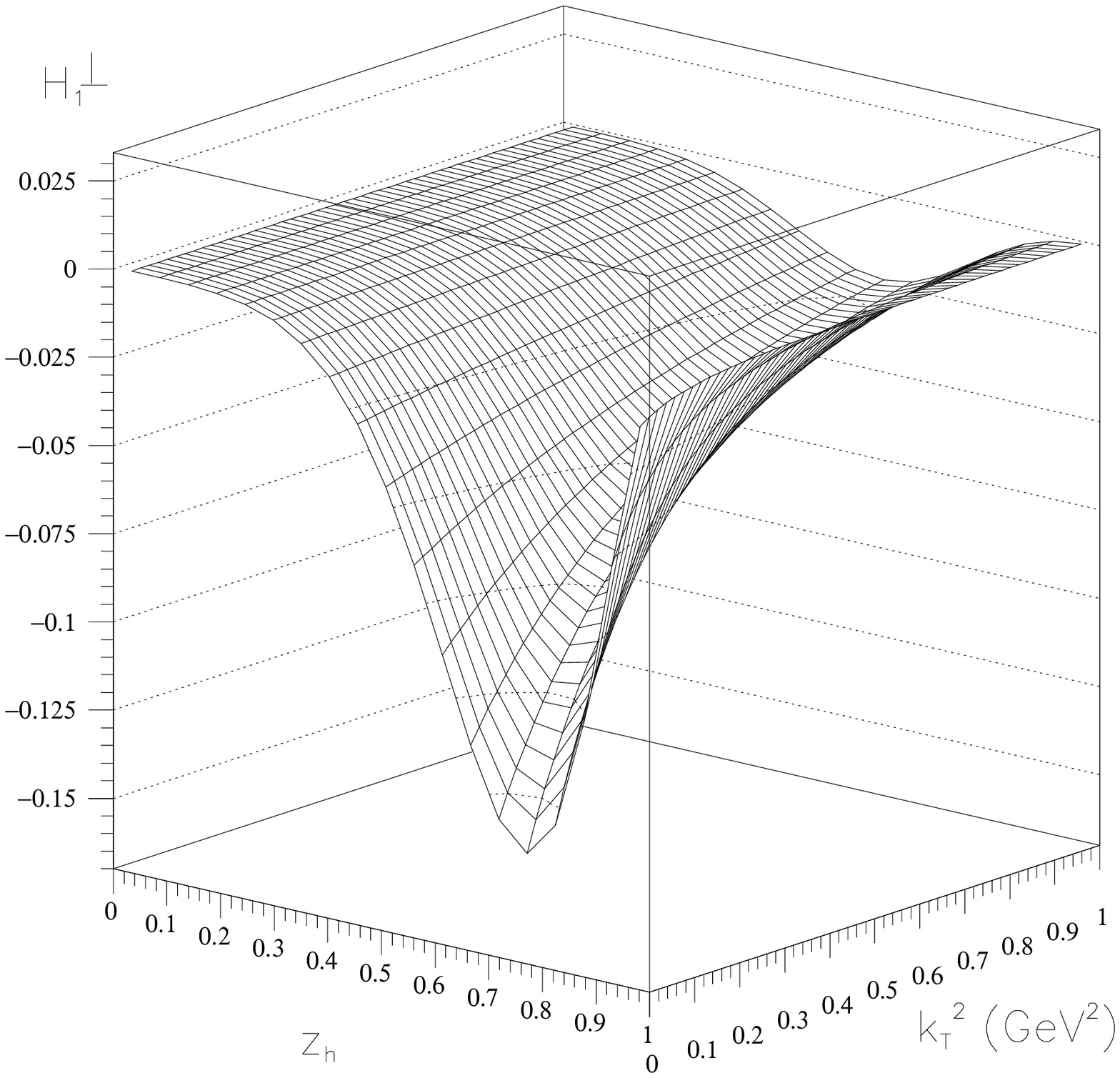} 
\end{center}
\vspace{3mm}
\begin{center} 
\includegraphics[width=55mm]{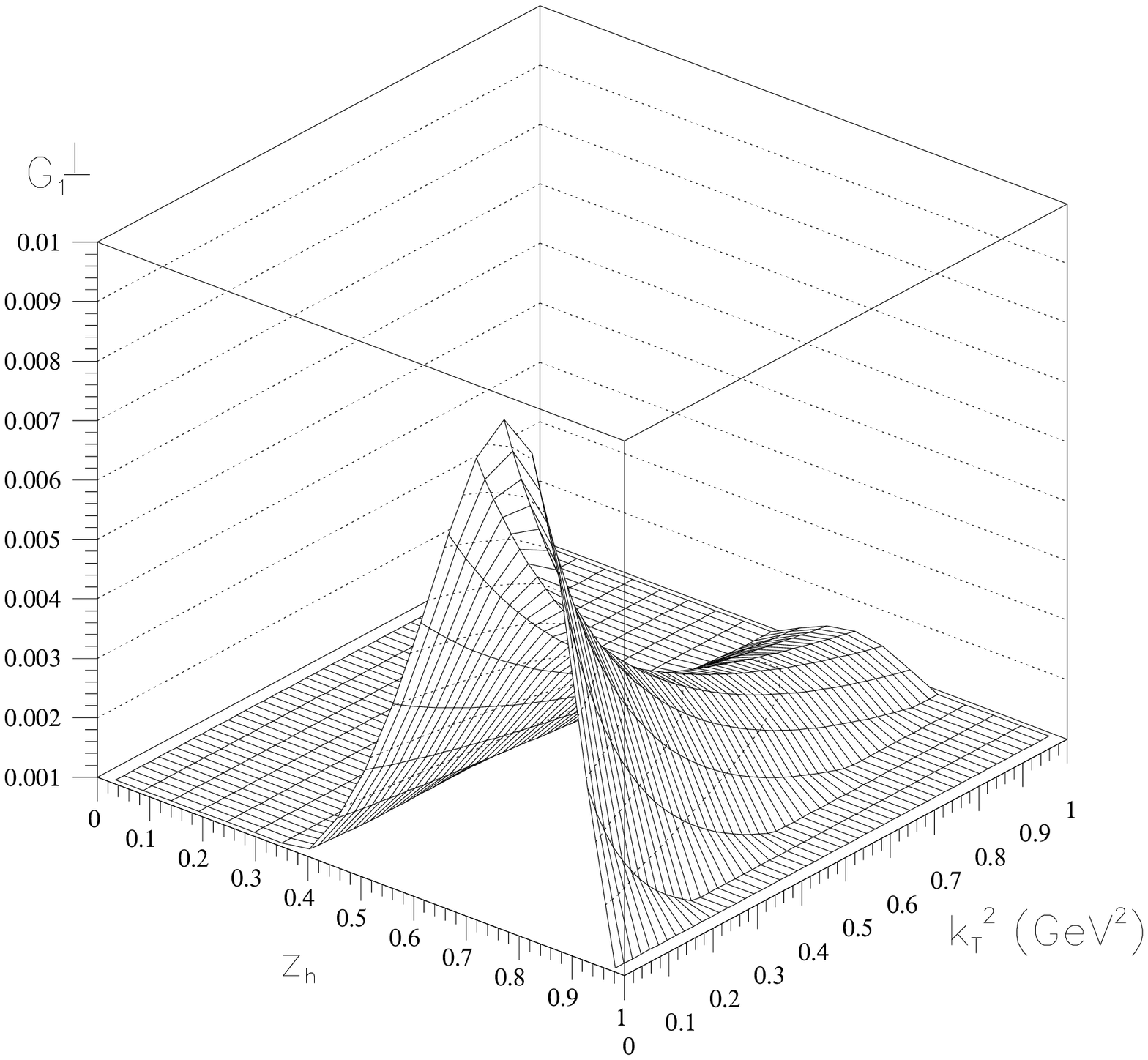}
\begin{minipage}[b]{55mm} 
The FF $H_1^{\newangle \; u\rightarrow p+\pi}$, 
$H_1^{\perp \; u\rightarrow p+\pi}$, 
$G_1^{\perp \; u\rightarrow p+\pi}$ are shown 
as functions of $z_h$, ${\vec k}_\sT^{\,2}$ at a fixed value 
$\xi=0.7$. Similar results are obtained for different values of $\xi$ in 
the kinematically
allowed range. The maximum sensitivity to the fragmentation mechanism is 
concentrated around the kinematical range where the pair takes roughly 80\% of 
the jet energy ($z_h \sim 0.8$) and has a small 
transverse momentum 
with respect to the jet axis ($\vec k_\sT^2 \lsim 0.4$ GeV$^2$).
\end{minipage} 
\end{center}

\section*{Acknowledgments}
This work is part of the TMR program ERB FMRX-CT96-0008. We thank 
P.J.~Mulders
for fruitful discussion.

\end{document}